\documentclass{article}

\usepackage{arxiv}
\usepackage[numbers,sort&compress]{natbib}

\usepackage[utf8]{inputenc} 
\usepackage[T1]{fontenc}    
\usepackage{hyperref}       
\usepackage{url}   
\usepackage{algorithm}
\usepackage{booktabs}       
\usepackage{amsfonts}       
\usepackage{nicefrac}       
\usepackage{microtype}      
\usepackage{lipsum}
\usepackage{graphicx}
\usepackage{amsmath} 
\usepackage{amssymb} 
\usepackage{caption}
\usepackage{pgfplots}
\usepackage{tikz}
\usepackage{enumitem}

\usetikzlibrary{positioning,calc,decorations.pathreplacing}

\usepackage{pgfplotstable}
\pgfplotsset{compat=1.18}

\usepackage[noend]{algpseudocode} 

\usepackage[utf8]{inputenc}

\usepackage{algpseudocode} 

\graphicspath{ {./images/} }

\title{Estimation of Phylogenetic Tree using Gene Sequencing Data \\[0.5em]
\large Published in 2019 4th International Conference on Electrical Information and Communication Technology (EICT)}

\author{
S M Rafiuddin \\
Department of Computer Science and Engineering \\
Bangladesh University of Engineering and Technology \\
\texttt{torifat.cs@gmail.com} \\
}

\begin{document}
\maketitle

\begin{abstract}
Phylogenetic tree is an important way in Bioinformatics to find the evolutionary relationship among biological species. In this research, a proposed model is described for the estimation of a phylogenetic tree for a given set of data. To estimate a phylogenetic tree there are certain necessary steps have to be considered. Gene sequences are useful data resources to estimate the relationship among species at the molecular level. In this research, the approach is to create a fusion between the existing models for the estimation of phylogenetic tree and a population-based meta-heuristics approach i.e. Genetic Algorithm. NCBI's Entrez databases are used for the acquisition of gene sequencing data. A modular parallel approach is applied to handle this dataset efficiently. This paper illustrates a proposed model to create an independent platform for phylogenetic tree estimation. Existing benchmark approaches for tree population that are used for population-based meta-heuristic search techniques to find the best possible phylogenetic tree estimation for a given set of gene sequence data.
\end{abstract}


\section{Introduction}

Phylogenetic Tree is an approach to represent the evolutionary relationship among biological species in an account of their similarities and dissimilarities on genetic characteristics. Phylogenetic tree is growing its importance enormously in the field of Bioinformatics, because there is a large field of genetic diversity and it is growing at a rapid speed. Evolutionary lineage is also estimated from phylogenetic trees. To make an estimation of Phylogenetic Tree, gene sequencing data is an important aspect. To represent multisequence alignment and to identify signatures of conservation, phylogenetic trees of sequences are important. It represents how organisms are related.

A phylogenetic tree can be rooted or non-rooted. In a rooted phylogenetic tree, a path from the root to a node represents an evolutionary path, i.e.\ it gives directionality to evolutionary time. On the other hand, a non-rooted tree specifies relationships among taxa, but it does not represent directionality information.

There are usually three types of methods for constructing phylogenetic trees based on the explicit model of evolution and character \cite{Lemey2009}—
\begin{enumerate}
  \item Maximum Likelihood Methods
  \item Maximum Parsimony Methods
  \item Pairwise Distance Methods
\end{enumerate}

There are two types of mutations in the gene sequence. In synchronous mutation, amino-acid do not change and in nonsynchronous mutation amino acid changes \cite{Lemey2009}. Recombination occurs when the parts of different DNA strands are merged into a single DNA strand \cite{Lemey2009}. To estimate a phylogenetic tree of a species and to find the evolution and relationship among genes there can have several procedures \cite{Lemey2009}. Morphological characters are used before to estimate phylogenetic tree, but in modern days the molecular data is the main interest for the estimation of the phylogenetic tree \cite{Liu2015}.

This research work is to estimate a phylogenetic tree for a species or genes using its gene sequencing data applying a metaheuristics algorithm. As phylogenetic tree estimation is an NP-hard problem and it can be solved in polynomial time, our approach is to create an estimation of a phylogenetic tree using meta-heuristics algorithm through the entire search space.

There are a few specific reasons behind of this research. Phylogenetic tree is an important way to enrich our understanding of how species evolved through the evolution process. It creates a true measurement of how the species have to classify and it helps us to choose the parameters of the classification procedure. A phylogenetic tree of gene sequences would answer various biological questions involving the genetic biodiversity of any species.

Phylogenetics is growing as an important sub-part of the Bioinformatics field. Also, this type of research would also create a great interest in forensic analysis. It could help us to find the origin of pathogens \cite{Vazquez1998}.

In a world of limited earth surface, there is huge biodiversity. We need to preserve their phylogeny identification. For future biodiversity and beyond, this genetic information needs to store in this very structure to predict any biological occurrences ever happened \cite{Soltis2003}.

We have the data of gene sequences, and we need to estimate the associated phylogenetic tree using gene sequencing data. The standard Genetic Algorithm is used to estimate the best possible phylogenetic tree.

\section{Literature Review}

Recent advancements of phylogenetics are phylogenetic tree analysis, parsimony, distance measurement, and likelihood estimation \cite{Yang2012}. Liang Liu et al. \cite{Liu2015} used gene-scale data for the estimation of phylogenetic tree. Various tools for implementing the functionalities of phylogenetics are used here.

Genetic distance is an important aspect of determining the phylogenetic tree of a particular species. If the ancestral sequence is divide into subsequence then it is possible to show the mutational and diverge sequence of the phylogenetic tree \cite{Lemey2009}. Neighborhood joining method is another way to achieve phylogenetic identity. In this method, evolution is occurred in every pair in joining, and the sum of all branch lengths is measured of that tree \cite{Bruno2000}.

Bayesian phylogenetic analysis is another approach to estimate the position of a species using phylogenetic trees. In this process, interference of phylogeny is determined based on the next stage probabilities of the DNA sequence. It creates a conditional probability to calculate the estimation of a phylogenetic tree using Bayes' theorem \cite{Huelsenbeck2001}.

One of the robust methods for the classic changes of phylogenetic analysis is species tree methods using the concatenation technique \cite{Liu2015}. To find the history of a species of tree life, genome-scale data gives unprecedented opportunities. In comparison with the polymerase chain reaction era, some ancient methods like evolutionary genomics and molecular ecology, are replaced by newly invented methods.

In recent years, a new method was introduced by Hey \& Nielsen, and others, which has been widely used. Bayesian formats have been updated to accommodate whole-genome data \cite{Hey2007,Machado2002,Wakeley1997}.

Generally, two models are used in the estimation process of a phylogenetic tree. They are coalescent and concatenation models. Coalescent model is now used over concatenation method for using free recombination among genes \cite{Liu2015}.

One of the simplest coalescent models that have recently been used is the multispecies coalescent model (MSC). It builds on the neutral coalescent model for a single species, such as the Wright–Fisher model \cite{Rannala2003}. In this model, each branch of species is treated as a single, neutral coalescent population, and going backward in time \cite{Degnan2009}.

\section{Proposed Model for Phylogenetic Tree Estimation}

As the problem definition is an NP‐hard problem, there is a huge search space to deal with. Our approach is to use the Genetic Algorithm (GA) for this solution procedure. The methodologies will use NCBI's Entrez databases for the Gene Sequencing database \cite{Maglott2011}. The solution procedure is as follows—

\begin{enumerate}
  \item Take Gene Sequencing data as the initial population.
  \item Retrieve the homologous sequence from the gene sequence data. Homologous gene sequence alignment is retrieved by log‐odds matrices \cite{Lemey2009}, which is defined by
    \begin{equation}
      S(i,j) \;=\; S(j,i) \;=\; \log \frac{q_{ij}}{p_i * p_j}
      \tag{1}
    \end{equation}
    Here, $q_{ij}$ = Probability of finding gene $i$ and gene $j$ aligned in the gene sequence.\\
    $p_i$ = Proportion of gene $i$ in sequence and it is the probability of finding $i$ in a random position of a gene sequence.\\
    $p_i * p_j$ = Probability of finding gene $i$ and gene $j$ aligned in random.
  \item Find the multiple sequence alignment from the homologous sequence.  
\begin{figure}[ht]
  \centering
  \includegraphics[width=0.5\textwidth]{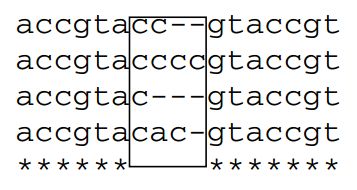}
  \caption{Multiple gene sequence alignment from homologous gene sequencing data.}
  \label{fig:msa}
\end{figure}

    Dynamic Programming (DP) is used to find the best alignment with associated scores. If two gene sequence $X$ and $Y$ are aligned in a 2D grid, $i$ is the index of sequence $X$ and $j$ is the index of sequence $Y$ then the best sub‐path will be determined by DP as \cite{Lemey2009}—
    \begin{equation}
      F(i,j) \;=\; \max\begin{cases}
        F(i-1,j-1) + s(X_i,Y_j)\\
        F(i-1,j) - g\\
        F(i,j-1) - g
      \end{cases}
      \tag{2}
    \end{equation}
    Here, $s(X_i,Y_j)$ is the score defined as the substitution matrix in the homologous gene sequence. As this formula works for only two sequences, this approach can be generalized for multiple sequence alignment for more than two sequences. These multiple generalized approaches would find the alignment that gives the best score for the following formula—
    \begin{equation}
      \sum_i \sum_j W_{ij} D_{ij}
      \tag{3}
    \end{equation}
    Here, $W_{ij}$ is the weight matrix and $D_{ij}$ is the distance matrix of $ij$ grid of multiple sequence alignment.  
    But if gaps of arbitrary size are possible to place in any particular position, then the affine gap penalty formula is used. By this technique, penalties are subtracted from the particular alignment score. Gap Penalties are determined by \cite{Lemey2009}—
    \begin{equation}
      \mathrm{GP} = g + e\,(l - 1)
      \tag{4}
    \end{equation}
    Here, $l$ = Length of the gap.\\
    $g$ = gap opening penalty, i.e.\ charged one for a gap.\\
    $e$ = gap extension penalty, i.e.\ charged once per start and end of the gap.
  \item Use the existing model for selecting the initial population. The existing models are as follows—
    \begin{description}
      \item[Maximum Likelihood Method:]  
        It is a probability‐based approach using a likelihood function. If the probability of $n$ tosses ends with $h$ head and the probability of the toss is $\Theta\in[0,1]$, then the likelihood function is defined as—
        \begin{equation}
          L(\Theta) = \Pr[H = h]
            = \binom{n}{h}\,\Theta^h\,(1-\Theta)^{n-h}
          \tag{5}
        \end{equation}
        Maximizing the likelihood function $L(\Theta)$, we can get—
        \begin{equation}
          L'(\Theta) = \frac{\partial \log[L(\Theta)]}{\partial \Theta}
            = \frac{h}{\Theta} \;-\; \frac{n-h}{1-\Theta}
          \tag{6}
        \end{equation}
        From the above theory, Maximum Likelihood Estimation (MLE) can be achieved for phylogenetic trees as—
        \begin{equation}
          L(\tau,\Theta) = \Pr(\mathrm{data}\mid \tau,\Theta)
          \tag{7}
        \end{equation}
        \begin{equation}
          \tau^{'}, \Theta^{'} = \arg\max_{\tau,\Theta} L(\tau,\Theta)
          \tag{8}
        \end{equation}
      \item[Maximum Parsimony Method:]  
        This approach for finding the optimal tree has two aspects \cite{Lemey2009}:  
        i) To determine the change of character and tree length.  
        ii) To search the entire space for all possible tree topologies to minimize the tree length.  
        If $\tau$ is the phylogenetic tree and $N$ is the number of characters, and $l_j$ is the length of a single sequence then the determined length is—
        \begin{equation}
          L(\tau) = \sum_{j=1}^{N} l_j
          \tag{9}
        \end{equation}
        If the nodes are terminal, the length of the binary phylogenetic tree is—
        \begin{equation}
          l_j = \sum_{k=1}^{2N-3} C_{a(k),b(k)}
          \tag{10}
        \end{equation}
        Here, $a(k)$ and $b(k)$ are states of the node of the branch $k$.
      \item[Pairwise Distance Method:]  
        The pairwise distance method is non-character-based and it is an explicit model of evolution. It has two steps \cite{Jukes1969}:  
        i) Evolutionary distances estimation. If $P$ is the percent difference between gene sequences $X$ and $Y$, then we get the genetic distance, $D$, as—
        \begin{equation}
          D = -\tfrac{3}{4}\ln\bigl(1 - \tfrac{4}{3}P\bigr)
          \tag{11}
        \end{equation}
        ii) By the calculated distance, infer the particular tree topology.
    \end{description}
\end{enumerate}

From these existing models described above, the phylogenetic tree population is produced, which will be used by the Genetic Algorithm by the following steps:

\begin{itemize}
  \item These initial populations chosen from the model are the input for the GA.
  \item Estimate the best tree using the Genetic Algorithm (GA).
\end{itemize}

\begin{algorithm}
\caption{Genetic Algorithm for Phylogenetic Tree Estimation}
\label{alg:ga}
\begin{algorithmic}[1]
\Require popsize \Comment{desired population size}
\State $P \gets \varnothing$
\For{$k = 1$ to popsize}
  \State $P \gets P \cup \{\text{new random individual from model selection outcome}\}$
\EndFor
\State $\mathrm{Best} \gets \varnothing$
\Repeat
  \ForAll{$P_i$ in $P$}
    \State AssessFitness($P_i$)
    \If{$\mathrm{Best} = \varnothing$ \textbf{or} Fitness($P_i$) $>$ Fitness($\mathrm{Best}$)}
      \State $\mathrm{Best} \gets P_i$
    \EndIf
  \EndFor
  \State $Q \gets \varnothing$
  \For{$k = 1$ to \(\frac{\text{popsize}}{2}\)}
    \State $P_a \gets$ SelectWithReplacement($P$)
    \State $P_b \gets$ SelectWithReplacement($P$)
    \State $(C_a, C_b) \gets$ Crossover(Copy($P_a$), Copy($P_b$))
    \State $Q \gets Q \cup \{\text{Mutate}(C_a), \text{Mutate}(C_b)\}$
  \EndFor
  \State $P \gets Q$
\Until{$\mathrm{Best}$ is the ideal solution or time runs out}
\State \Return $\mathrm{Best}$
\end{algorithmic}
\end{algorithm}

\begin{figure}[ht]
  \centering
  \includegraphics[width=0.8\textwidth]{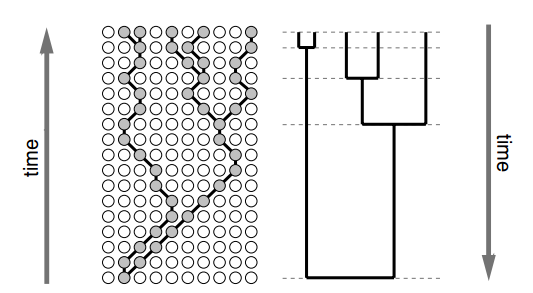}
  \caption{Phylogenetic tree estimation using the Genetic Algorithm.}
  \label{fig:ga-estimation}
\end{figure}

III. Assess fitness from the estimated tree using fitness function.
\begin{equation}
  E = \sum_{i=0}^{T-1} \sum_{j=i+1}^{T} W_{ij} \lvert d_{ij} - p_{ij} \rvert^{\alpha}
  \tag{12}
\end{equation}
Here, $E$ = Error of fitting estimates and the tree.  
$T$ = Number of taxa.  
$d_{ij}$ = Distance between taxa $i$ and $j$.  
$p_{ij}$ = Path length between taxa $i$ and $j$.  
$W_{ij}$ = Weight to separate the taxa $i$ and $j$.  
$\alpha$ = Arbitrary value between 1 and 2.

IV. Estimate best tree with measures of support.

e) Test hypothesis result.

This proposed solution is depicted in a diagram below—

\begin{figure}[ht]
  \centering
  \includegraphics[width=0.8\textwidth]{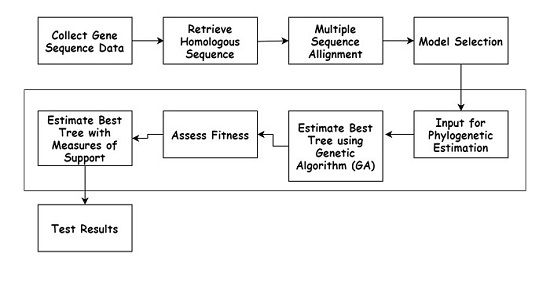}
  \caption{Diagram of the proposed system for the estimation of phylogenetic trees.}
  \label{fig:proposed-system}
\end{figure}

\section{Experiments and Results}

\subsection{Dataset Description}
The experiment of this research project dataset is NCBI's Entrez databases \cite{Maglott2011}. This database is a complete collection of information on molecular biology. This database is publicly available and directly downloadable for all kinds of users \cite{Lemey2009}. There are nine important NCBI dataset accessible via Geneious.

From these nine database Nucleotides, Unigene, Gene, Popset and Taxonomy dataset are important for estimation of phylogenetic tree for a selected species \cite{Lemey2009}. Nucleotide database contains non-redundant DNA sequence \cite{Lemey2009}. Unigene is a collection of GenBank sequences important for phylogenetic information grouped by gene \cite{Lemey2009}. Gene database is a collection of information about genetic loci \cite{Lemey2009}. PopSet is another kind of dataset included in NCBI's Entrez databases which is important for the estimation of phylogenetic trees because it contains the information of multiple sequence alignments of DNA sequences of different populations of species \cite{Lemey2009}. Taxonomy contains the information of all species, sub-species and higher and lower taxon and their position order for which there is at least one DNA sequence in databases \cite{Lemey2009}.

NCBI's Entrez database size is huge. Reading the entire dataset at a time is not efficient for a single machine. So, the database should parse before use. For example, only Human genome has a size of 116576 kB. This database can be converted to 6.1 GB file [17].

\subsection{Experimental Setup}
\textbf{Hardware Setup:}\\
Processor: AMD X4 Phenom II 965 BE 3.4 GHz\\
RAM: 8 GB DDR3 1333 MHz\\
HDD: 1 TB

\textbf{Software Setup:}\\
Operating System: Linux Mint XFCE 18.2\\
Programming Language: Python 3.4.1\\
Used packages: Biopython

\subsection{Results}

\begin{table}[ht]
\centering
\caption{Result Analysis of Phylogenetic Tree Estimation}
\label{tab:results1}
\begin{tabular}{r r r r r}
\hline
Number of & Origin of & Number of & Number of & Number of \\
gene sequences & Sequence & tree & tree generated & tree generated \\
per process & in Database & generated using & using Maximum & using \\
 &  & Maximum Likelihood Methods & Parsimony Methods & Pairwise Distance Methods \\
\hline
500   &   1 &  34 &  41 &  32 \\
1000  & 501 &  56 &  53 &  45 \\
2000  & 1501 & 59 &  62 &  69 \\
3500  & 1251 & 65 &  71 &  74 \\
5500  & 651 &  78 &  92 &  78 \\
\hline
\end{tabular}
\end{table}

\begin{table}[ht]
\centering
\caption{Result Analysis of Error Estimation of Phylogenetic Tree}
\label{tab:results2}
\begin{tabular}{r r}
\hline
Number of population used & Error (\%) after GA \\
for Genetic Algorithm & estimates best possible \\
Search Space & phylogenetic trees \\
\hline
12   & 8.70 \\
50   & 7.84 \\
91   & 7.63 \\
201  & 6.17 \\
310  & 5.89 \\
\hline
\end{tabular}
\end{table}

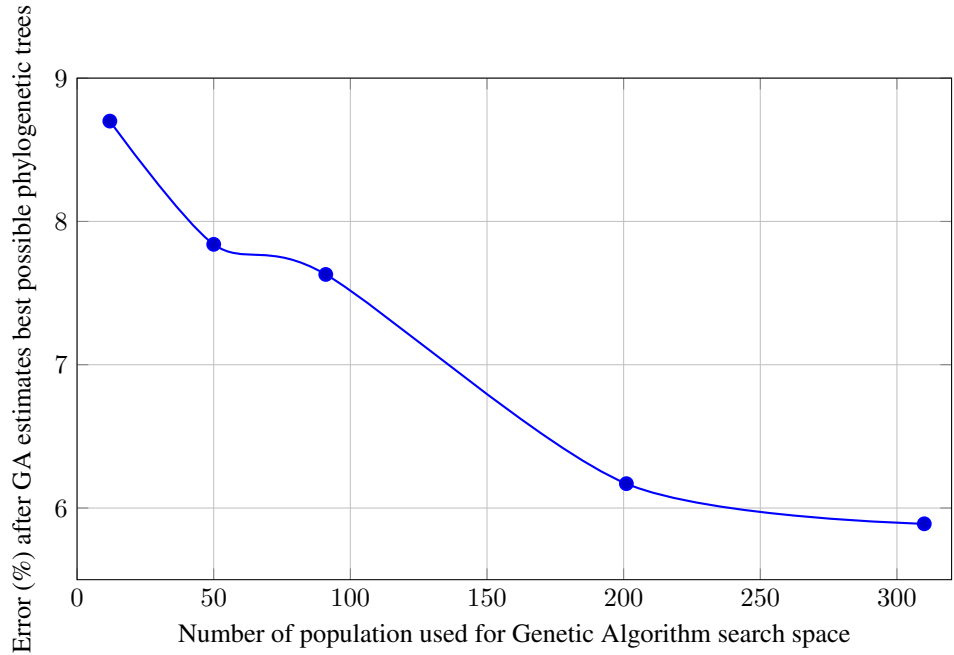
\begin{figure}[ht]
\centering
\begin{tikzpicture}
  \begin{axis}[
    width=0.8\textwidth,
    height=0.5\textwidth,
    xlabel={Number of population used for Genetic Algorithm search space},
    ylabel={Error (\%) after GA estimates best possible phylogenetic trees},
    xmin=0, xmax=320,
    ymin=5.5, ymax=9.0,
    xtick={0,50,100,150,200,250,300},
    ytick={6,7,8,9},
    grid=both,
    grid style={line width=0.1pt, draw=gray!20},
    major grid style={line width=0.2pt, draw=gray!50},
    every axis plot/.append style={
      thick,
      mark=*,
      mark options={fill=white, scale=1.2},
    },
    smooth
  ]
    \addplot coordinates {
      (12,8.70)
      (50,7.84)
      (91,7.63)
      (201,6.17)
      (310,5.89)
    };
  \end{axis}
\end{tikzpicture}
\caption{Error (\%) after GA estimates best possible phylogenetic trees versus number of population used for Genetic Algorithm search space.}
\label{fig:error}
\end{figure}

\section{Conclusion}

This research has the goal of finding the best possible phylogenetic tree minimizing the error rate. The construction of phylogenetic trees depends on slightly how the gene sequence datasets are organized as well as the model is used for the estimation process. After generating the phylogenetic trees from different models, the trees are applied through the Genetic Algorithm using population genetics to find the best possible phylogenetic tree for a particular species. This proposed system for estimating phylogenetic trees shows that using meta-heuristic approaches for a certain search space can come up with a better phylogenetic tree with a minimum error rate. However, the more gene sequences it gets, the more processing time it needs to estimate the possible phylogenetic tree. So, it needs a higher processing power to deal with a large gene sequence database to estimate a phylogenetic tree for a particular species. To know the evolutionary relationship of a species, phylogenetic tree estimation will grow more attention in the future research field of bioinformatics. The future work of this research is to optimize the approach up to a certain level to cope with processing time and minimize the error rate.  

\section*{Acknowledgment}

The author(s) would like to thank Professor Dr. Mohammad Sohel Rahman, Department of Computer Science and Engineering (CSE), Bangladesh University of Engineering and Technology (BUET), for his kind support and enthusiasm. Also, this research work is very much grateful to Pavel Pevzner, Distinguished Professor, Department of Computer Science, University of California, San Diego, and Phillip Campeau, Assistant Teaching Professor, Department of Computational Biology, Carnegie Mellon University, for their MOOC course on Bioinformatics in Coursera \cite{BioinformaticsMOOC}.

\end{document}